\documentstyle[12pt]{article}
\newcommand{\be}{\begin{equation}}
\newcommand{\ee}{\end{equation}}
\newcommand{\bea}{\begin{eqnarray}}
\newcommand{\eea}{\end{eqnarray}}
\newcommand{\beas}{\begin{eqnarray*}}
\newcommand{\eeas}{\end{eqnarray*}}
\newcommand{\bi}{\begin{itemize}}
\newcommand{\ei}{\end{itemize}}
\newcommand{\bc}{\begin{center}}
\newcommand{\ec}{\end{center}}
\newcommand{\bfl}{\begin{flushleft}}
\newcommand{\efl}{\end{flushleft}}
\newcommand{\bfr}{\begin{flushright}}
\newcommand{\efr}{\end{flushright}}
\newcommand{\f}{\frac}

\def\6{\partial} \def\a{\alpha} \def\b{\beta}

  \def\th{\theta}
  \def\l{\lambda}
\def\m{\mu} \def\n{\nu}  
 \def\s{\sigma} \def\t{\tau}

  \def\O{\Omega}

\newcommand{\HH}{{\cal H}}

\newcommand{\ZZ}{{\cal Z}}
\begin{document}

\title{Entropy of Bosonic Open String and Boundary Conditions}

\author{M. C. B. Abdalla$^1$\thanks{Email: {\tt mabdalla@ift.unesp.br}}
E. L. Gra\c{c}a$^2$$^3$\thanks{Email: {\tt edison@cbpf.br}}
I. V. Vancea$^1$$^2$\thanks{Email: {\tt ivancea@ift.unesp.br}}
\\
$^1$ Instituto de F\'{\i}sica Te\'{o}rica, Universidade 
Estadual Paulista\\
Rua Pamplona 145, 01405-900 S\~{a}o Paulo-SP, Brasil\\
$^2$ Centro Brasileiro de Pesquisas F\'{\i}sicas\\
Rua Dr. Xavier Sigaud 150, CEP 22290-180 Rio de Janeiro RJ, Brasil\\
$^3$ Departamento de F\'{\i}sica \\
Universidade Federal Rural do Rio de Janeiro \\
BR 465 - 07 - Serop\'{e}dica CEP 23890-000 RJ, Brasil}

\maketitle

\begin{abstract}
The entropy of the states associated to the solutions of the equations of 
motion of the bosonic 
open string with combinations of Neumann and Dirichlet boundary conditions 
is given. Also,
the entropy of the string in the states 
$\left| A^i \right> = \alpha^i_{-1}\left|
0 \right>$ and $\left| \phi^a \right> = \alpha^a_{-1}\left| 0 \right>$ that 
describe the massless fields on the world-volume of the $Dp$-brane is computed. 
\end{abstract}

\section{Introduction}

Recently, there has been some interest in formulating the $D$-branes at 
finite temperature for several reasons.
On one 
hand, the relation between the $D$-branes and the field theory at 
finite temperature represents an interesting problem by itself which could
help us to get a better understanding of the physical properties of 
$D$-branes. Some progress in this direction has been 
done in the low energy limit of string theory where the $D$-branes are 
solitonic solutions of (super)gravity. In this limit, the 
thermodynamics of $D$-branes has been formulated in the framework of 
(real time) path-integral formulation of field theory at finite temperature 
\cite{mvm,od,rsz,kog1,kog2,kog3,rab1,rab2,gub}
On the other hand, one would like to 
understand the statistical properties of some systems that can be 
described in terms of $D$-branes and anti-$D$-branes as, for example, the 
extreme, near-extreme 
and Schwarzschild black-holes 
\cite{as,sv,ls,hp,bfks1,cm,ks,eh,dmkr,hm,ml,bfk,lt,dm,ast,ivvol,ms,dgk}.

In the other well understdood 
limit of string theory, the perturbative limit, the geometric 
information about the $D$-branes is lost. In that case, the 
$D$-branes are more appropriately described by a superposition of coherent 
states in the Fock space of closed string sector 
\cite{gw,bg,gg,ml1,ck,dv1,dv2} that should satisfy a certain set of 
boundary conditions. They correspond to a combination of Dirichlet and Neumann
boundary conditions which should be imposed at the endpoints of the 
open string in the open string channel. The intuitive interpretation of
$D$-branes as coherent boundary states is maintained at finite temperature
if one formulates the problem in the framework of the Thermo Field Dynamics 
(TFD). The reason for that is that the thermal dependence is implemented 
through the so called thermal Bogoliubov operators \cite{tu} which
preserve the form of the relations at zero temperature. Working with TFD 
instead of real-time path integral formalism represent just a convenient
way of formulating the problem since it is known that the two formalisms are 
equivalent at the thermodynamical equilibrium. The main idea of
TFD is to interpret the statistical average of any quantity $Q$ over a 
statistical ensemble as the expectation value of $Q$ in a thermal vacuum 
\be Z^{-1}(\b_T ){\mbox Tr}[Q e^{-\b_T H}]~=~
\left\langle\! \left\langle 0 (\b_T ) \left| Q \right| 0 (\b_T )\right\rangle 
\!\right\rangle ,
\label{tfd}
\ee
where $\b_T ~=~(k_B T)^{-1}$ and $k_B$ is the Boltzmann's constant \cite{tu}.
The thermal vacuum is a vector in the direct product space of the original
Fock space by an identical copy of it. The second copy of the
Fock space provide the necessary states to describe the thermal effects. 

The TFD was used to discuss the ideal gas of strings, to construct a bosonic 
open string field theory at finite temperature \cite{yl1,yl2,yl3} and to
prove its renormalizability \cite{fn1,hf,fn2}. These studies were motivated
by the need of understanding string cosmology and the ensembles of 
strings in general. However, when one applies the TFD to $D$-branes one has 
to take some
care \cite{cav,ivv,cav1,cav2}. Indeed, from the beginning, the $D$-branes are 
defined as states in 
the Fock space of the first quantized single string. The conformal field 
theory that underlies the construction describes
the bosonic vacuum of string theory. When passing to finite temperature,
one interprets the thermal string as a model for {\em the thermal excitations 
of the bosonic vacuum} of string theory. Therefore, the TFD is applied to the
conformal field theory in two dimensions.
Thus, the thermal $D$-branes should be interpreted as a coherent  
boundary states in the Fock space of the thermal excitations.
Another remark is that according to (\ref{tfd}), the entropy of the system is 
defined as the
thermal vacuum expectation value of the entropy operator. However, our basic 
system is the string and we can compute the entropy of the thermal string
excitations by applying (\ref{tfd}). What about the entropy of the $D$-brane
states? At the moment, there is no known theory in which the $D$-branes are
described by vacuum-like states or by states created from a vacuum. Thus,
in order to compute the entropy of $D$-brane states, we have to calculate
the average value of the entropy operator of the bosonic string in the thermal
$D$-brane state. These two remarks make our TFD approach to the thermal 
strings different from the previous ones. 

The point of view adopted in \cite{cav,ivv,cav1,cav2} is that of $D$-branes
as states in the closed string sector.  The TFD is applied to it. 
However, for the sake of completness, one should construct the thermal theory
for the open string sector as well. In this case, one can wonder if the 
boundary conditions have any influence on the thermal properties of the system.
The entropy of the open string in its termal vacuum is given by the sum over
all space-time directions of the entropy of the scalar fields and thus it 
does not depend on the boundary conditions. However, if the string is in
a state that depends on the boundary conditions, the entropy computed as
the average of the entropy operator in that state depends on them. Since the
bosonic vacuum depends on what sort of boundary conditions we impose to the
equations of motion (for example, the Dirichelt boundary conditions will 
indicate the presence of branes), we would like to look to the entropy in the
states associated to the general solutions of the equations of motion with
a mixture of Dirichlet and Neumann boundary conditions.
The aim of the present paper is to present the TFD construction for the
bosonic string in the sense explained above, i. e. for bosonic string vacuum,
and to compute its entropy in states with explicit dependence on boundary
conditions. In particular, we calculate the entropy
of open strings in the thermal massless states which can be interpreted as
thermal massless fields living on the $D$-brane.
There are at least two reasons for which such of computation should be 
performed. Firstly, it represents a natural step in developing a theory
of string vacua and thermal branes in TFD approach. In this context, the states
that present an explicit dependence on the boundary states play a fundamental
role. Secondly, it might help to understand the entropy of the $D$-branes
in the perturbative limit of string theory.

\section{The Bosonic String in TFD}

Consider a bosonic open string in flat spacetime. In order to avoid the 
introduction of ghosts, we choose to work in the light-cone gauge
$X^{0} \pm X^{9}$ so that the spacetime indices run over the set
$\m , \n = 1,2,\cdots , 24$. The string can be quantized in the canonical
formalism by interpreting the most general solution of the equations of
motion as an operator on the Fock space and by imposing the canonical 
commutation relations. The dependence on the world-sheet parameters 
$\left( \t , \s \right)$ is determined by the boundary conditions which can
be of the Neumann (N) or of the Dirichlet type, independently at each end of 
the string. Let us consider the string with NN boundary conditions. The other
cases are treated in exactly the same manner.
The general solutions $X^{\m}(\t , \s )$ of the equations of motion of
the bosonic open string with the Neumann boundary conditions at the two ends 
are giving by the following relations
\be 
X^{\m}(\t , \s )~=~ x^{\m} + 2 \a '\t p^{\m} + i \sqrt{2\a '} \sum_{n \neq 0} 
\f{\a^{\m}_n}{n}e^{-in\t} \cos n\s . 
\label{solNN}   
\ee
Here, $x^{\m}$ and $p^{\m}$ are the the canonical coordinates and momenta of 
the center of mass of the string.
It is useful to introduce the oscillator operators
\be
A^{\m}_n = \f{1}{\sqrt{n}}\a^{\m}_n ~~~;~~~A^{\m \dagger}_n = 
\f{1}{\sqrt{n}}\a^{\m}_{-n}~~,
~~~ n > 0,
\label{oscop}
\ee 
that satisfy the canonical commutation relations among themselves and commute 
with the coordinates and momenta of the center of mass. The vacuum is defined 
to be invariant under translations
\be
\left| \O \right\rangle  ~= ~\left| 0 \right\rangle  \left| p \right\rangle ,
\label{vac}
\ee 
where
\bea
A^{\m}_n \left| 0 \right\rangle ~ &=& ~ 0 ~~,~~~ \forall n, \label{vacosc}\\
\hat{p}^{\m} \left| p \right\rangle ~ & = & ~ p^{\m} \left| p \right\rangle 
\label{vacmom}.
\eea

Now let us construct the thermal bosonic string. According to TFD, firstly 
one has to duplicate the system. The identical copy of it, denoted by 
$~\tilde{}~$, must be independent of the original string 
\cite{cav,ivv,cav1,tu}. The Hilbert space of the total system is the tensor 
product of the two Hilbert spaces $\HH$ and $\tilde{\HH}$. The operators for 
the two strings commute among themselves. By duplicating the system, new
degrees of freedom are associated to the string. They provide enough room to 
accomodate the thermal properties of the bosonic vacuum of string. To this end,
the new copy of the string cannot represent a physical system. Otherwise, the
tilde Hilbert space will describe just the dynamical degrees of freedom of the
second string, and no degrees of freedom will be left for the thermal effects.
\footnote{We would like to thank to A. L. Gadelha for illuminating discussions
on this point.}
They can be implemented in the enlarged Hilbert space by constructing a map
called the thermal Bogoliubov transformation
that maps the vacuum of the string to the thermal vacuum defined by 
(\ref{tfd}). 
It turns out that there are several possibilities to construct 
such of map and that the corresponding operators generate a thermal $SU(1,1)$
group. But if we stick to unitary and tilde invariant transformations we are
left with just a single operator in the present case \cite{cav2,um}.
The Bogoliubov transformation is defined for
each oscillator $n$ in each direction $\m$ by the following relation
\be
G^{\m}_n ~=~-i \th_n(\b_T)(A_n \cdot \tilde{A}_n - \tilde{A}_n^{\dagger} \cdot
 A_n^{\dagger} ).
\label{bogolop}
\ee  
Here, $\th_n(\b_T)$ is a parameter related to the Bose-Einstein distribution 
of the oscillator
$n$. The dot in 
(\ref{bogolop}) represents the Euclidean scalar product in the target space.
The vacuum of the oscillators at
$T \neq 0$ is given by the following relation
\be
\left. \left| 0(\b_T ) \right\rangle \! \right\rangle ~= ~\prod_{ m > 0} 
e^{-iG_m} 
\left. \left| 0 \right\rangle \! \right\rangle ,
\label{vacT}
\ee
where $\left. \left| 0 \right\rangle \! \right\rangle= 
\left| 0 \right\rangle\tilde{\left| 0 \right\rangle} $ and
\be
G_n = \sum_{\m=1}^{24}G^{\m}_{n}.
\label{opbogoltot}
\ee
The total vacuum at $T\neq 0$ is constructed in the same way as (\ref{vac}), 
i.e.
\be
\left. \left| \O (\b_T) \right\rangle \! \right\rangle ~=~ 
\left. \left| 0(\b_T) \right\rangle \! \right\rangle
\left| p \right\rangle \left| \tilde{p} \right\rangle. 
\label{totvacT}
\ee
The state (\ref{vacT}) is called the thermal vacuum since it is annihilated by
all thermal annihilation operators constructed by acting with the 
Bogoliubov operators (\ref{opbogoltot}) onto the oscillator operators of string
\be
A^{\m}_{n}(\b_T) ~= ~ e^{-iG_n}A ^{\m}_{n}e^{iG_n}~~~,~~~
\tilde{A}^{\m}_{n}(\b_T) ~= ~ e^{-iG_n}\tilde{A} ^{\m}_{n}e^{iG_n}.
\label{annihT}
\ee
The same action of the Bogoliubov operators gives any other thermal operator.
In particular, one constructs the creation operators 
$A^{\m \dagger}_{n}(\b_T)$ and $\tilde{A}^{\m \dagger}_{n}(\b_T)$ 
which, together with (\ref{annihT}) define the set of thermal string 
oscillators since they satisfy the same commutation relations as the operators
at $T=0$. The coordinates and the momenta of the center of mass of string are 
invariant under the Bogoliubov map. Thus, one can construct the string 
solution $X^{\m}(\b_T)$ at $T \neq 0$ by replacing 
the operators in (\ref{solNN}) with the corresponding operators at $T \neq 0$. 
In exactly the same way one can construct the generators of the Virasoro 
algebra in terms of thermal oscillator operators and show that 
the conformal symmetry of the solution is preserved. Consequently, the
Bogoliubov transformation maps the two string copies into two thermal string 
copies \cite{ivv,cav1}. However, the interpretation in terms of original and 
tilde string modes is lost at non-zero temperature since the Bogoliubov 
operators (\ref{bogolop}) mix the two copies of strings. Also, the thermal
vacuum should be invariant under the tilde operation and therefore the tilde
and non-tilde zero-temperature excitations are generated simultaneously at 
$T \neq 0$. This is an intrinsic fact of the TFD construction. Heuristically,
one would say that the thermal string excitations, being a mixture of 
zero-temperature string and tilde string excitations, carry thermal degrees of
freedom beside dynamical degrees of freedom (in zero-temperature language) 
which is in agreement with our intuition.

\section{The Entropy in States with Boundary Dependence}

Since we are treating the string as a set of bosonic oscillators,
the entropy operators for the bosonic open string follow 
directly from the definition  \cite{tu}
\bea
K ~&=& ~\sum_{\m = 1}^{24}\sum_{n=1}^{\infty}( A^{\m \dagger}_n A^{\m}_n 
\log \sinh^2 \th_n -A^{\m}_n A^{\m \dagger}_n \log \cosh^2 \th_n )
\label{entropy}\\
\tilde{K} ~&=&  \sum_{\m = 1}^{24}\sum_{n=1}^{\infty}(\tilde{A}^{\m \dagger}_n
\tilde{A}^{\m}_n \log \sinh^2 \th_n -
\tilde{A}^{\m}_n \tilde{A}^{\m \dagger}_n \log \cosh^2 \th_n ).
\label{entropytilde}
\eea
The vacuum of the present theory is invariant under the tilde operation and
all the physical information is contained in the operators without 
tilde. Therefore, we are going to calculate the matrix elements only of the
entropy (\ref{entropy}). The basic idea is to compute firstly the matrix 
elements to the contribution to the entropy of the excitations in each 
direction of space-time. Thus, it is useful to write (\ref{entropy}) in the 
form
\be
K ~ =~ \sum_{\m = 1}^{24} K^{\m},
\label{decomentrop}
\ee
where the notation is obvious. 

The entropy of the string, as given by the formula (\ref{tfd}), represents the
sum of the entropy of all oscillators and it is straight-forward to write it
down. Next, we would like to know the entropy of the string in states 
associated to the most general solution of the equation of motion. This is a
similar situation to computing the entropy of $D$-branes. 
In this case, the entropy is a 
world-sheet function and its dependence on the world-sheet parameters is
dictated by the boundary conditions imposed on the string equations of motion.
Upon quantization, the general solution of the string equations of motion with
given boundary conditions at the two ends becomes an operator acting on the
thermal Fock space. The state 
$\left.\left| X^{\m}(\b_T) \right\rangle\!\right\rangle$ obtained by applying
(\ref{solNN}) on the thermal vacuum (\ref{totvacT}) can be associated to it.
Let us work out the matrix elements 
$\left\langle\!\left\langle X^{\m}(\b_T)\left| K^{\rho} \right|
 X^{\m}(\b_T )
\right\rangle\!\right\rangle$ for the NN boundary conditions in some detail.
The computations envolving the other boundary conditions 
follow the same line.

The matrix element  
$\left\langle\!\left\langle X^{\m}(\b_T)\left| K^{\rho} \right| X^{\m}(\b_T )
\right\rangle\!\right\rangle$
can be split in two parts: one part containing only the oscillator 
contribution and a part that 
contains the coordinates and the momenta of the center of mass
\bea
\left\langle \!\left\langle X^{\mu}(\b_T)\left| K^{\rho} \right| 
X^{\m}(\b_T )\right\rangle\!\right\rangle
~&=&~{\mbox{CM terms}} \nonumber\\
&-& 2\a ' \sum_{n,k,l >0}\f{e^{i(l-n)\t}}{\sqrt{ln}}\cos n\s \cos l\s
\left[ (T_1)^{\m\rho\n}_{nkl} + (T_2)^{\m\rho\n}_{nkl}\right],
\label{matrentr}
\eea
where
\bea
(T_1)^{\m\rho\n}_{nkl} ~&=&~
\left\langle \tilde{0} \left| \left\langle 1^{\m}_n \left|
\prod_{m>0}e^{-iG_m}A^{\rho \dagger}_{k}A^{\rho}_{k}\log \sinh^2 \theta_k 
\prod_{s>0}e^{iG_s}
\right| 1^{\n}_l \right\rangle \right| \tilde{0} \right\rangle
\left\langle \tilde{p} \left|  \tilde{q} \right\rangle \right.
\left\langle p \left| q \right \rangle \right. ,\nonumber\\
(T_2)^{\m\rho\n}_{nkl} ~&=&~
- \left\langle \tilde{0} \left| \left\langle 1^{\m}_n \left|
\prod_{m>0}e^{-iG_m}A^{\rho }_{k}A^{\rho \dagger}_{k}\log \cosh^2 \theta_k 
\prod_{s>0}e^{iG_s}
\right| 1^{\n}_l \right\rangle \right| \tilde{0} \right\rangle
\left\langle \tilde{p} \left|  \tilde{q} \right\rangle \right.
\left\langle p \left| q \right \rangle \right.
\label{Ts}
\eea
and
\be
\left| 1^{\m}_{l} \right\rangle ~= ~A^{\m \dagger}_{l} \left| 0 \right\rangle.
\label{field}
\ee
We consider the usual normalization of the momentum states in a finite volume 
$V_{24}$ in the transverse space
\bea
\left.\left\langle p \right| q \right\rangle &~=~& 2 \pi 
\delta^{(24)} (p-q) \label{normstate1}\\
(2\pi )^{24}\delta^{(24)}(0) &~=~& V_{24}.
\label{normstate2}
\eea
It follows from a simple algebra that the pure oscillator contribution to the
matrix element of the entropy operator has the following form 
\bea
\left\langle\!\left\langle X^{\m}(\b_T)\left| K^{\rho} \right| X^{\m}(\b_T )
\right\rangle\!\right\rangle
~&=&~{\mbox{CM terms}} 
-2 \a ' (2 \pi)^{(48)} \delta^{\m \n} \delta^{(24)}(p-q)\delta^{(24)}
(\tilde{p}-\tilde{q}) \times 
\nonumber\\
& &\sum_{n>0}\f{1}{n}\cos^2 n\s [\log (\tanh \th_n )^2\delta^{\rho \n } -
\delta^{\rho \rho }\sum_{k>0} \delta_{k k}].
\label{matrelemCM}
\eea

The terms containing the center of mass can be divided at their turn in to two 
sets: the ones containing only the coordinates and momenta operators and the 
ones containing the 
contributions from the 
oscillators. The terms in coordinates and momenta can be calculated by using 
the completness relation of the eigenstates of momenta operators 
$\left| x \right\rangle$ together with the coordinate-momenta matrix 
\be
\left. \left\langle x \right| p \right\rangle ~=~(2\pi \hbar ) ^{-12}
e^{i p \cdot x / \hbar}.
\label{matrixxp}
\ee
In order to compute the terms containing oscillator contribution one can 
either express the oscillator operators at zero temperature in terms of 
operators at $T \neq 0$ or write the vacuum at $ T \neq 0 $ in terms of 
vacuum at zero temperature. The two ways lead to the same result, however in 
the first case one has to deal only with polynomial relations in creation and 
annihilation  operators at finite temperature.  Using the properties 
of the Bogoliubov operators deduced in \cite{ivv,cav1} and after a  simple 
algebra one can show that the contribution coming from terms that mix the 
center of mass operators with the oscillator part cancells. The non-vanishing 
terms are all proportional to 
$\left\langle \!\left \langle 0(\b_T) \left| K^{\rho} \right| 0(\b_T) 
\right\rangle \!\right\rangle$ which represents the entropy of an infinite 
number of bosonic oscillators in the $\rho$'th directions of space-time. 
Putting all the results together we obtain the following relation 
\bea
& &\left\langle \!\left\langle X^{\mu}(\b_T)\left| K^{\rho} \right| 
X^{\m}(\b_T )\right\rangle\!\right\rangle
~=~ \nonumber\\
&- &(2\pi \hbar)^{-24}\left[ (2\pi\hbar)^{24}(2\a ' \t)^2 p^{\m} p'^{\n}
\delta^{(24)}(p-p') 
%\right.
%\nonumber\\ 
+ 
%\left.
2\a ' \t (I^{\m}_2p'^{\n} + I'^{\n}_2 p^{\m}) + 
I^{\m}_2I^{\n}_2\prod_{j \neq \m , \n}I^j_1 \right]\nonumber\\ 
&\times &\delta^{(24)}(\tilde{p} - \tilde{p'})\sum_{m=1} 
\left[{\mbox n}^{\rho}_m 
\log {\mbox  n}^{\rho}_m + (1- {\mbox  n}^{\rho}_m) 
\log( 1- {\mbox  n}^{\rho}_m ) 
\right]
%\nonumber\\
- 2 \a ' (2 \pi)^{(48)} \delta^{\m \n} \delta^{(24)}(p-p')\nonumber\\
&\times &\delta^{(24)}(\tilde{p}-\tilde{p'})  
%\nonumber\\
\sum_{n>0}\f{1}{n}\cos^2 n\s \left[ \log (\tanh \th_n )^2\delta^{\rho \n } -
\delta^{\rho \rho }\sum_{k>0} \delta_{k k} \right],
\label{entropyRho}
\eea
where the unidimensional integrals on the finite domains 
$ x \in \left[ x_0, x_1\right] $ are
given by
\bea
I_1 &~=~& -i \hbar (p'-p)^{-1}\left[ e^{\f{i}{\hbar}(p'-p)x_1} - 
e^{\f{i}{\hbar}(p'-p)x_0}\right]
\label{int1}\\
I_2 &~=~& -i \hbar (p'-p)^{-1}\left[  -i\hbar I_1 + 
x_1e^{\f{i}{\hbar}(p'-p)x_1} 
- x_0 e^{\f{i}{\hbar}(p'-p)x_0}\right].
\label{int2}
\eea 
Here, the momentum of the final and initial states are denoted by 
$\left| p \right\rangle$ and
$\left| p' \right\rangle$, respectively and 
\be
{\mbox n}^{\rho}_{m} ~=~ \left\langle \!\left\langle 0(\b_T) \left| 
A^{\rho \dagger}_m A^{\rho}_m \right| 0(\b_T) \right\rangle\!\right\rangle 
= \sinh^2 \th_m \label{Nnumber} 
\ee 
represents the number of string excitations in the thermal vacuum.

Relation (\ref{entropyRho}) represents the matrix element of the entropy 
operator $K^{\rho}$ 
between two general states described by solutions of the equations of motion 
of the open bosonic string with NN boundary conditions. The total energy is a 
sum over all directions in the transverse space. 
Since the effect of the boundary condition appears only in the 
dependence on the world-sheet 
parameters, it is easy to write down the similar relations for the other 
boundary conditions DD, DN and ND. In these cases, there are no operators 
associated with the coordinate and momenta of the center of mass of the open 
string but rather constant position vectors associated to the endpoints of 
string. Therefore, there is no contributions from these terms to the entropy. 
The terms that mix the coordinates of the endpoints with the oscillators 
vanish for the same reason as in the NN case. The only non-vanishing terms in 
the matrix elements of the entropy are 
\bea 
{\mbox DD}&:& 
\left\langle \!\left\langle X^{\mu}(\b_T)\left| K^{\rho} \right| 
X^{\m}(\b_T )\right\rangle\!\right\rangle
=
%\nonumber\\ 
2 \a ' (2 \pi)^{(48)} \delta^{\m \n} \delta^{(24)}(p-p')\delta^{(24)}
(\tilde{p}-\tilde{p'})  
\nonumber\\
&\times &
\sum_{n>0}\f{1}{n}\sin^2 n\s \left[ \log (\tanh \th_n)^2\delta^{\rho \n } -
\delta^{\rho \rho }\sum_{k>0} \delta_{k k} \right]
\label{entrDD}\\
{\mbox DN}&:&
\left\langle \!\left\langle X^{\mu}(\b_T)\left| K^{\rho} \right| 
X^{\m}(\b_T )\right\rangle\!\right\rangle
= 
%\nonumber\\
2 \a ' (2 \pi)^{(48)} \delta^{\m \n} \delta^{(24)}(p-p')\delta^{(24)}
(\tilde{p}-\tilde{p'})\!  
\nonumber\\
&\times &
\sum_{r=\ZZ + 1/2 }\!\f{1}{r}\sin^2 r\s \left[ \log (\tanh \th_r)^2
\delta^{\rho \n } -
\delta^{\rho \rho }\sum_{k>0} \delta_{k k} \right]
\label{entrDN}\\
{\mbox ND}&:&
\left\langle \!\left\langle X^{\mu}(\b_T)\left| K^{\rho} \right| 
X^{\m}(\b_T )\right\rangle\!\right\rangle
= 
%\nonumber\\
2 \a ' (2 \pi)^{(48)} \delta^{\m \n} \delta^{(24)}(p-p')
\delta^{(24)}(\tilde{p}-\tilde{p'}) 
\nonumber\\
&\times & \!\sum_{r=\ZZ + 1/2}\!\f{1}{r}\cos^2 r\s \left[ \log (\tanh \th_r)^2
\delta^{\rho \n } -
\delta^{\rho \rho }\sum_{k>0} \delta_{k k} \right]
\label{entrND}
\eea
Here, $\ZZ + 1/2$ are the half-integer numbers. The contribution of just a 
single field is obtained by taking $\m = \rho = \n$.
The relations (\ref{entropyRho}), (\ref{entrDD}), (\ref{entrDN}) and 
(\ref{entrND}) represent the entropy of the states associated to the general
solutions of the equations of motion. They give the entropy as a function on 
the world-sheet, therefore it should not be taken for the entropy of the 
bosonic string vacuum which is simply the sum over all space-time directions
of the entropy of massless scalar bosons and it does not depend on the
boundary conditions as it should. 

\section{Discussions}

The matrix elements (\ref{entropyRho}), (\ref{entrDD}), (\ref{entrDN}) and 
(\ref{entrND}) can be used to calculate the entropy of various states 
of open strings with different boundary conditions. As an example, 
consider 
the massless states of strings ending on one $Dp$-brane. It is 
known that these states describe massless fields on the world-volume of 
the brane. They form an $U(1)$ multiplet 
$A^j ~=~\a^{j}_{-1}\left| 0 \right\rangle$, where $j=1, 2, \ldots, p$ and 
a set of $(24-p)$ scalars $\phi^a ~=~\a^{a}_{-1} \left| 0 \right\rangle$ 
where $a=p+1, \ldots, 24$. The contribution to the entropy of the 
corresponding string states with DD, DN and ND boundary condition can be 
computed by truncating (\ref{entrDD}), (\ref{entrDN}) and (\ref{entrND}) 
to the first oscillator term or computing the matrix elements of $K^\rho$ 
in the thermal states 
\be 
\left.\left| \Psi^{\l} (\b_T) \right\rangle \!\right\rangle 
~=~\a^{\l}_{-1}(\b_T )\left. \left| 0 (\b_T ) \right\rangle \!\right\rangle
\label{fieldsatT}
\ee
multiplied by the function on $\t$ and $\s$ that is obtained 
from the corresponding boundary conditions. 
A simple algebra gives the follwing expression for the entropies of the 
$U(1)$ multiplet and the set of scalars, respectively 
\bea 
E_{ \{ A \} } &~=~& 2\a ' p \left( - \sin^2 \s \sum_{n=1}\log \cosh^2 \th_n 
+ 4 \cos^2 \s \sum_{r \in \ZZ+ 1/2} \log \cosh^2 \th_r \right),
\label{entru1}\\
E_{ \{ \phi \} } &~=~& 2\a ' (24-p) \left( - \sin^2 \s \sum_{n=1}\log \cosh^2 
\th_n + 4 \cos^2 \s \sum_{r \in \ZZ+ 1/2} \log \cosh^2 \th_r \right)
\label{entrscal}.
\eea
In (\ref{entru1}) and (\ref{entrscal}) the first term represents the 
contribution from the DD sector while the second one is obtained from
DN and ND sectors. 

In conclusion, we have calculated the entropy of string in states that 
depend explicitely on the boundary
conditions. To this end, we have 
picked up the most general solution of the equations of motion with 
all possible combinations of Dirichlet and Neumann boundary conditions and 
we computed the matrix elements of the entropy in the state associated to it. 
The relevant relations 
are (\ref{entropyRho}), (\ref{entrDD}), (\ref{entrDN}) and (\ref{entrND}). 
It is interesting to observe that only the NN entropy depends on $\hbar$. 
In the semiclassical limit $\hbar \rightarrow 0$ the pure momentum 
contribution becomes irrelevant. The dominating term is the same that 
dominates the infinite tension limit $\a' \rightarrow 0$. In this case the 
entropy of the DD, DN and ND strings vanishes. 

In particular, we have applied the general results to the computation of the 
entropy of the massless string states that generate the field theories on the 
world-volume of a $Dp$-brane. In the case of a stack of $N$ parallel D-branes 
the gauge multiplet is of $U(N)$ and in order to find the total 
entropy of it we just have to sum over the internal indices. However, the 
scalars become non-commutative and the present approach should be extended 
to include non-commutative fields \cite{caei}.
Constructing the $D$-branes in the thermal string field theory represents a 
different but interesting topic.
If such of theory were given, one could get
insights in the problem of the entropy of the $D$-brane which, in the
present formulation, seems to suffer from certain anomalies \cite{cav1}.

{\bf Acknowledgements}
We would like to thank N. Berkovits, A. L. Gadelha, P. K. Panda,
B. M. Pimentel, H. Q. Placido and W. P. de Souza for useful discussions. 
I. V. V. also acknowledge to S. A. Dias and 
J. A. Helay\"{e}l-Neto for hospitality at DCP-CBPF and GFT-UCP 
where part of this work was done. I. V. V. was supported by a FAPESP postdoc 
fellowship.

\end{document}